\documentstyle[preprint,eqsecnum,aps,psfig]{revtex}
\tightenlines
\def\beq{\begin{equation}}
\def\eeq{\end{equation}}
\def\bea{\begin{eqnarray}}
\def\eea{\end{eqnarray}}
\def\nnu{\nonumber}
\def\tst{\textstyle}

\def\fno#1{Fig.~\ref{#1}}
\def\cno#1{\cite{#1}}
\def\eno#1{Eq.~(\ref{#1})}
\def\etwo#1#2{Eqs.~(\ref{#1}) and (\ref{#2})}
\def\rno#1{Ref.~\cite{#1}}
\def\Slab#1{Sec.~\ref{#1}}

\def\al{\alpha}
\def\be{\beta}

\def\dta{\delta}
\def\eps{\epsilon}

\def\kap{\kappa}
\def\lam{\lambda}
\def\sig{\sigma}
\def\om{\omega}

\def\Gam{\Gamma}
\def\Dta{\Delta}
\def\Lam{\Lambda}

\def\apx{\approx}
\def\ptl{\partial}

\def\hf{{1\over2}}
\def\tshf{\tst\hf}
\def\quar{{1\over 4}}
\def\tsqua{\tst\quar}

\def\lp{\left(}
\def\rp{\right)}

\def\ham{{\cal H}}
\def\ket#1{|#1\rangle}

\def\tran#1#2{\langle#1|#2\rangle}
\def\avg#1{\langle#1\rangle}
\def\mel#1#2#3{\langle#1|#2|#3\rangle}

\def\bH{{\bf H}}
\def\bJ{{\bf J}}

\def\bx{{\bf x}}

\def\bz{{\bf z}}

\def\xhat{\bf{\hat x}}
\def\zhat{\bf{\hat z}}

\def\Fe8{Fe$_8$}
\def\hsc{{\ham_{\rm sc}}}

\def\baj{\bar J}

\def\baj{\bar J}

\def\Mpm{M_{\pm}}
\def\ompm{\om_{0\pm}}
\def\omzm{\om_{0-}}
\def\omzp{\om_{0+}}
\def\mpb{m'_b}
\def\nph{{\nu' + \tshf}}
\def\lph{{\ell' + \tshf}}
\def\mzm{m_{0-}}

\def\Dll{\Dta({\ell',\ell''})}

\begin{document}
\draft

\title{Quenched Spin Tunneling and Diabolical Points
in Magnetic Molecules: II. Asymmetric Configurations}

\author{Anupam Garg$^*$}
\address{Department of Physics and Astronomy, Northwestern University,
Evanston, Illinois 60208}

\date{\today}

\maketitle

\begin{abstract}
The perfect quenching of spin tunneling first predicted for a
model with biaxial symmetry, and recently observed in the magnetic
molecule \Fe8, is further studied using the discrete phase integral
(or Wentzel-Kramers-Brillouin) method. The analysis of the previous
paper is extended to the case where the magnetic field has both hard
and easy components, so that the Hamiltonian has no obvious symmetry.
Herring's formula is now inapplicable, so the problem is solved by
finding the wavefunction and using connection formulas at every
turning point. A general formula for the energy surface in the vicinity
of the diabolo is obtained in this way. This formula gives the tunneling
apmplitude between two wells unrelated by symmetry in terms of a small
number of action integrals, and appears to be generally valid, even for
problems where the recursion contains more than five terms.
Explicit results are obtained
for the diabolical points in the model for \Fe8. These results exactly
parallel the experimental observations. It is found that the leading
semiclassical results for the diabolical points appear to be exact,
and the points themselves lie on a perfect centered rectangular lattice
in the magnetic field space. A variety of evidence in favor of this
perfect lattice hypothesis is presented.
\end{abstract}

\widetext

\section{Introduction}
\label{intro}
\subsection{The story so far}
\label{sofar}

In a previous paper\cite{numeins}, hereafter cited as I, we studied
the tunneling of a spin governed by the Hamiltonian
\beq
\ham  = -k_2 J_z^2 + (k_1 - k_2) J_x^2 - g\mu_B \bJ\cdot\bH, \label{ham}
\eeq
where $\bJ$ is a dimensionless spin operator, and $k_1 > k_2 > 0$.
This Hamiltonian is the simplest descriptor of the magnetic properties
of the molecule [(tacn)$_6$Fe$_8$O$_2$(OH)$_{12}$]$^{8+}$ (or just \Fe8
for short), with $J=10$, $k_1 \apx 0.33$~K, and $k_2 \apx 0.22$~K
\cite{cs,alb,rc}. Interest in this molecule arises because of its
rich low temperature magnetic behavior, which include hysteresis at
the level of one molecule \cite{cs}, and more recently, the
discovery \cite{ws} of an entire lattice of diabolical
points \cite{hlh,bw} in its magnetic spectrum, a subset of which was
predicted to exist earlier \cite{ag1}. It is the latter property
that we wish to continue investigating in this paper.

In paper I, attention was confined to the case where the external
magnetic field $\bH$ is along the hard axis $\bx$. In this case,
the classical energy, which may be viewed as the expectation value
$\avg\ham$ of the Hamiltonian in a spin-coherent state, is symmetric
about the $xy$ plane, and the
problem is analogous to that of a massive particle in one dimension
in a reflection symmetric double well potential. It is then natural
to consider the tunneling between the symmetrically related states
localized in the left- and right-hand wells. In the magnetic case,
the analogous states are those with predominatly positive and
negative values of $J_z$, at least as long as $H_x$ is not so large
as to bring the classical minima very close to the $x$ axis.
The surprise is that the tunnel splitting between the ground states
oscillates as a function of $H_x$, vanishing exactly at a series of
points. In fact, the splitting between higher pairs of levels also
vanishes at just these points, as noted in I and earlier studies
\cite{vf,agletts}.

In addition to observing the quenching of ground state tunneling
when $\bH\|\bx$, however, Wernsdorfer and Sessoli \cite{ws} also
performed experiments with $H_z \not= 0$. The reflection symmetry
of the classical energy is now lost, but if the value of $H_z$
is chosen properly, it is possible to bring an excited state in
the positive $J_z$ well into approximate degeneracy with the
ground state of the negative $J_z$ well. The new discovery by
them is that if $H_x$ is now varied, the tunnel splitting between
the degenerate or quasi-degenerate levels again oscillates. 
It is theoretically understood that if both $H_z$ and $H_x$ are
properly tuned, the splitting vanishes exactly in this case
too \cite{vf,aglt}. Experimentally, of course, one can never
see a perfect zero in the splitting, and Wernsdorfer and Sessoli
only see a minimum in the Landau-Zener-St\"uckelberg transition
rate between the levels in question. The minima are so deep,
however, that there is little doubt that the underlying tunneling
matrix element is quenched.

When $\bH\|\bx$, the Hamiltonian (\ref{ham}) is invariant under
a 180$^\circ$ rotation about $\bx$, and the quenchings can be
understood from the viewpoint of the von Neumann-Wigner theorem as
allowed crossings of energy levels with different parities under
this rotation. A similar argument can be made when
$\bH\|\bz$. When $\bH$ has both $x$ and $z$ components, however,
the Hamiltonian has no obvious symmetry, and the above theorem
states that an intersection of two energy levels is infinitely
unlikely as a single
parameter in the Hamiltonian is varied. For a real symmetric
Hamiltonian, it is known \cite{va} that one must vary two
parameters to obtain an intersection. Since we can choose the
matrices of both $J_z$ and $J_x$ to be real in the $J_z$ basis,
these two parameters can be taken as $H_x$ and $H_z$. The isolated
points \cno{fn1} in the $H_x$-$H_z$ plane where any two levels intersect
are precisely what Herzberg and Longuet-Higgins \cite{hlh} call
conical intersections and what Berry and Wilkinson \cite{bw} call
diabolical points. The latter terminology originates in the
resemblance of the energy surface---a double cone with a common
vertex \cno{et}---to an Italian toy called {\it the diavolo}.

\subsection{Content and plan of this paper; the perfect lattice
            hypothesis}
\label{thispap}

In this paper we shall allow $\bH$ to lie in the $x-z$ plane,
with a view to studying the tunneling in the asymmetrical well,
and locating the diabolical points. As in paper I, our analysis
is based on the discrete phase integral (DPI), or
Wentzel-Kramers-Brillouin method \cite{dpirefs}. This method
is semiclassical in character, with $1/J$ playing the same role as
$\hbar$ in the continuum phase integral method. 
To introduce this method, let us write the Schr\"odinger equation
$\ham\ket{\psi} = E\ket{\psi}$ in the $J_z$ eigenbasis
$\{\ket m\}$. With
$J_z \ket m = m\ket m$, $\tran{m}{\psi} = C_m$,
$\mel{m}{\ham}{m} = w_m$, and $\mel{m}{\ham}{m'} = t_{m,m'}$ ($m \ne m'$),
we have
\beq
\mathop{{\sum}'}_{n = m-2}^{m+2} t_{m,n} C_n + w_m C_m = E C_m, \label{Seq}
\eeq
where the prime on the sum indicates omission of the $n=m$ term.
A vivid picture of the approximation can be obtained if we think of
\eno{Seq} as the tight-binding model for an electron in a one-dimensional
lattice with sites labelled by $m$, and on-site ($w_m$) and hopping
($t_{m,m\pm1}$, $t_{m,m\pm2}$) energies. If $J \gg 1$, these matrix
elements vary slowly with $m$, on a length scale of order $J$ in fact.
We may then use the approximation of semiclassical dynamics by working
entirely in terms of wavepackets whose spatial extent is much less than
the length scale over which the properties of the lattice vary, i.e., $J$,
and whose spread in Bloch vectors is much less than the width of the
Brillouin zone, i.e., $2\pi$. These ideas have close counterparts in the
continuum quasiclassical method, and the DPI method is nothing but the
discrete analog. 

When $\bH\|\bx$, the problem can also be approached using
instantons---indeed the oscillations in the splitting were discovered
in this way. When $\bH$ has other components besides $H_x$, however,
the DPI method is, to our knowledge, the only successful one to date.
Villain and Fort's approach \cite{vf} is also an approximate version
of this method that makes artful use of some special features of the
\Fe8 problem and works for small values of the field. Our analysis
is more prosaic, and almost self-evident
once one has understood how to deal with the new feature in
\eno{Seq}---the presence of second neighbour hopping. This gives rise
to novel turning points with no continuum analogues. Connection formulas
for these turning points are given in \rno{jmp2}. We have quoted the
results of our analysis before \cite{aglt,agletts}(b), but the details
are only being presented here.

Our main result for the specific Hamiltonian (\ref{ham})
is for the locations of the diabolical points. We find that 
the $\ell'$th level in the negative $J_z$ well (where $\ell'=0$
denotes the lowest level) and the $\ell''$th level in the positive
$J_z$ well are degenerate when (see \fno{dpts})
\bea
{H_z(\ell',\ell'') \over H_c}
    &=& {\sqrt{\lam}(\ell'' - \ell') \over 2 J}
                 \label{plhz} \\
{H_x(\ell',\ell'') \over H_c}
   &=& {\sqrt{1-\lam} \over J}
        \left[ J - n - \tshf (\ell' + \ell'' + 1) \right],
                 \label{plhx}
\eea
with $n = 0, 1, \ldots, 2J - (\ell' + \ell'' + 1)$. Here,
$\lam = k_2/k_1$, and $H_c = 2 k_1 J/g\mu_B$.

It should be stressed that \etwo{plhz}{plhx} only represent the first
terms of a series in $1/J$, and that our DPI calculations
give no reason to believe that the higher order terms are absent.
Yet a large amount of empirical evidence suggests just this, i.e.,
that the {\it formulas are exact as written}! That the diabolical
points lie on a perfect centered rectangular lattice, and that many
pairs of levels are simultaneously degenerate, we refer to as
the perfect lattice hypothesis. It has previously been made for the
ground state on the $H_z = 0$ line by Villain and Fort \cno{vf},
and extended to include simultaneous degeneracy of the higher states
by us \cite{agletts}(b). This was shown by perturbation theory
in $\lam$ to $O(\lam^3)$ for all $J$, and analytically for 
$J \le 2$. We have since found analytically that for $J = 3/2$,
the additional diabolical point at
$(H_x,H_z) = (\sqrt{1-\lam},\sqrt\lam)H_c/3$ is exact. Of course,
Kramers theorem guarantees double degeneracy of {\it all} energy
eigenvalues at the point $H_x = H_z = 0$ for all half-integral $J$.
We have also checked this hypothesis numerically for a variety of
values of $\lam$, and  $J$ up to 10. We present our results for
$J =5$ in Table I. The deviations in the locations of the diabolical
points from the perfect lattice hypothesis predictions are never more
than $10^{-10}$, certainly much less than order $1/J$. In fact,
the values listed are below the numerical tolerance that we prescribed.

Further support for the perfect lattice hypothesis comes from the
fact that it is consistent with the following
duality property of the Hamiltonian (\ref{ham}). If we set $k_1 =1$,
$\ham$ may be written as $\ham(\lam,H_x,H_z)$, showing
its dependence on the three parameters, $\lam$, $H_x$ and $H_z$. By a
$90^{\circ}$ rotation about the axis $(\xhat + \zhat)/\sqrt{2}$, we
obtain the transformation
\beq
\ham(\lam,H_x,H_z) \leftrightarrow -\ham(1-\lam,H_z,H_x).
                         \label{dual}
\eeq
In particular, the spectra of the two Hamiltonians are so related, and
ranking the levels is order of increasing energy, we see that
if the levels with ordinal numbers $k$ and $k+1$ are degenerate when
$H_x = f_x(\lam)$ and $H_y = f_y(\lam)$, then level numbers
$2J+2 -k$ and $2J+1-k$ are degenerate when $H_x = f_y(1-\lam)$
and $H_y = f_x(1-\lam)$. These conditions do not constrain the
functions $f_x$ and $f_y$ in any real way, however, so the discovery
of formulas (\ref{plhz}) and (\ref{plhx}) must presently be put down
to serendipity.

The plan of our paper is as follows. In \Slab{lowdown}, we analyze
the asymmetric double well problem in completely general terms. We
briefly review the DPI method and the asymmetric double-well problem
in one dimensional quantum mechanics. In contrast to the situation
that prevails in that case, we must deal with
{\it four} independent DPI wavefunctions at every point. Consequently
the matching problem is harder and more subtle. Its study forms the
bulk (subsections C to G) of the analysis. The main result of this
analysis is a formula for the diabolo [\eno{diab}], and a two level
system Hamiltonian [\eno{htls}]. The tunneling amplitude appearing
in these formulas is found in a general form [\eno{defDta}] involving
a small number of action integrals running between various turning
points. Readers who are not interested in the details of the analysis
should skip to subsection H and read on from \eno{diab}. They may also
find the last paragraph of \Slab{lowdown} interesting. 

In \Slab{appfe8}, we apply the general analysis to the model
(\ref{ham}) for \Fe8. To keep the analysis tractable, we will assume
that $H_z/H_c$ is small, although there is no reason why one could
not apply the results of \Slab{lowdown} numerically for arbitrary
values of $H_z$. It may in fact be interesting to do this to further
investigate the perfect lattice hypothesis.  Further, as discussed in
\Slab{appfe8}, the duality property of $\ham$ immediately extends our
work to large $H_z$ and small $H_x$.

\section{General Formula for Diabolo in terms of Action Integrals}
\label{lowdown}
\subsection{Summary of DPI method, critical curves, etc.}
\label{recap}

Very briefly, the DPI analysis proceeds as follows. (See \rno{jmp2} and
paper I for details.) Let the energy
of an electron wavepacket in the equivalent tight-binding model be given
by $\hsc(q,m)$ where $q$ and $m$ are the mean wavevector and position
of the wavepacket, respectively. Holding $m$ fixed, we may think of
$\hsc(q,m)$ as a dispersion relation for the local energy band. The
mean velocity of the wavepacket is then given by
$v(q,m) = \ptl\hsc(q,m)/\ptl q$. If we regard $m$ as a continuous
variable, and approximate $w_m$, $t_{m,m\pm1}$, and $t_{m,m\pm 2}$ by
smooth functions $w(m)$, $t_1(m)$, and $t_2(m)$, then
\bea
\hsc(q,m) &=& w(m) + 2t_1(m) \cos q + 2t_2(m) \cos(2q),
          \label{hsc} \\
v(q,m) &=& -2\sin q(m) \bigl(t_1(m) + 4 t_2(m) \cos q(m)\bigr).
     \label{vm}
\eea
In close analogy with the continuum method, one can show that the
quasiclassical solution to the wavefunction for a given energy $E$ is
given by linear combinations of solutions of the form
\beq
C_m \sim {1 \over \sqrt{v(m)}}\exp\lp i\int^m q(m') dm'\rp,
    \label{Cwkb}
\eeq
where $q(m)$ and $v(m)$ are given by
\beq
E = \hsc\bigl(q(m),m\bigr),
    \quad v(m) = v\bigl(q(m),m\bigr). \label{Eandv}
\eeq

It follows from \eno{Cwkb} that the quasiclassical solution breaks
down when $v(m) = 0$---which we call a turning point, as in the continuum
case. At these points, the solutions must be augmented by connection
formulas. In addition to the values $q=0$ and $q=\pi$, $v(m)$ also
vanishes at $q=q^*(m) = \cos^{-1}\bigl(-t_1(m)/4t_2(m)\bigr)$. 
We will get turning points whenever $E$ equals $U_0(m)$, $U_{\pi}(m)$,
or $U_*(m)$, where these three functions are $\hsc(0,m)$,
$\hsc(\pi,m)$, and $\hsc(q^*,m)$ respectively. These three curves,
which we call {\it critical curves}, collectively play the same role
as the potential energy in the continuum case. 

For \Fe8, in the same reduced variables as in I [$\mu = m/\baj$,
energies in units of $k_1\baj^2$, $\baj = (J+\tshf)$], the on-site
and hopping energies are given by
\bea
w(m) &=& \tshf (1+\lam)(1-\mu^2) - 2 h_z \mu, \label{wm} \\
t_1(m) &=& - h_x (1-\mu^2)^{1/2}, \label{t1m}\\
t_2(m) &=& \tsqua (1-\lam) (1-\mu^2).
\eea
Here, $\lam = k_2/k_1$, $h_x = JH_x/\baj H_c$, $h_z = JH_z/\baj H_c$,
and $H_c = 2k_1 J/g\mu_B$. Thus the critical curves are given by
\bea
U_0(\mu) &=& 1-\mu^2 - 2 h_x (1-\mu^2)^{1/2} - 2h_z\mu,
     \label{U0} \\
U_{\pi}(\mu) &=& 1-\mu^2 + 2 h_x (1-\mu^2)^{1/2} - 2h_z\mu,
     \label{Upi} \\
U_*(\mu) &=& \lam (1-\mu^2) - {h_x^2 \over 1-\lam}
        - 2h_z\mu. \label{Ustar}
\eea
These curves are shown in \fno{fig2}.
The energy $U_{\pi}(m)$ is the upper edge of the band $\hsc(q,m)$ for
all $m$. The lower band edge is given by
$U_0(m)$ for values of $m$ close to $\pm J$, and by $U_*(m)$ for $m$
in the central region. In the central region, the second neighbour
hopping element $t_2(m)$ is sufficiently large that the local energy band
$\hsc(q,m)$ has its global minimum not at $q=0$, but at $q^*$. Since
the energy $U_0$ lies in the band, $U_{\pi} > U_0 > U_*$. By
contrast, in the outer $m$ regions, $\hsc(q,m)$ has only one minimum (at
$q=0$) and only one maximum (at $q=\pi$) for real $q$. The energy $U_*$
lies outside the band, so that once again $U_{\pi} > U_0 > U_*$. The
curves $U_0(m)$ and $U_*(m)$ touch one another with a common tangent at
$m = \pm m^*$, where
\beq
m^* = \baj \left[ 1 - {h_x^2 \over (1-\lam)^2} \right]^{1/2}.
             \label{mustar}
\eeq

The turning point where $E = U_*$ is special if it happens at a
value of $m$ where $U_*$ lies below the lower band edge. The wavevector
$q^*$ at the turning point is then complex, and the wavefunction
$C_m$ changes from a decaying (or growing) exponential on one side
to a decaying (or growing) exponential with an oscillatory envelope
on the other side. These turning points are the new feature caused by
second neightbor hopping that we referred to in \Slab{intro}.

\subsection{Nature of asymmetric double well wavefunctions}
\label{psigrows}
To understand how the DPI solutions are to be used to find the
eigenstates, it is useful to discuss the corresponding problem for
an asymmetric double well in the continuum case. Suppose the potential
is as drawn in \fno{fig3}. The potential minima are at $x_{0\pm}$, and
for the energy $E$ drawn, the turning points are at $x'_a$ and $x'_b$
in the left well, and $x''_b$ and $x''_a$ in the right well. We use the
quasiclassical approximation to find a wavefunction $\psi'(x)$ on the
left hand side as follows. First, we choose the solution which decays
exponentially as $x - x'_a\to - \infty$. This solution is matched via
connection formulas at $x'_a$ to an oscillatory solution in the region
$x'_a < x < x'_b$. We then use connection formulas at $x'_b$ to find
the quasiclassical solution in the region $x > x'_b$. We repeat this
procedure on the right hand side to find a wavefunction $\psi''(x)$
that decays to zero as $x-x''_a \to \infty$. The last step is to
demand that the wavefunctions $\psi'(x)$ and $\psi''(x)$ be the same
in the central region, i.e., in the vicinity of $x=0$. This demand
will be unsatisfiable for an arbitrarily chosen energy $E$, and will
provide one with the eignevalue condition.

There are two remarks that we wish make about the above procedure.
The first remark concerns the basic nature of the solution. 
In general, in the central region, the left solution $\psi'(x)$
will be a linear combination of two parts, $\psi'_d(x)$
and $\psi'_g(x)$, that are exponentially decaying and growing as
$x-x'_b$ increases, respectively. Likewise the right solution,
$\psi''$, will be a sum of parts that decay ($\psi''_d$) and
grow ($\psi''_g$) as $x''_b - x$ increases. The key point
is that the growing parts $\psi'_g$ and $\psi''_g$ must be present
in an eigenstate, for without them, there is no way that the values
and slopes of $\psi'$ and $\psi''$ could be made to agree at $x=0$,
say.

The second remark is technical. If the potential well is reasonably
parabolic near the minimum at $x_{0-}$, the Schr\"odinger equation
can be solved directly for any choice of $E$ in terms of parabolic
cylinder functions, and we can always find a linear combination that
will decay to zero as $x - x'_a \to -\infty$. This linear combination
will have both growing and decaying pieces as $x - x'_b$ grows. In
this way we can obtain the wavefucnction $\psi'(x)$ on the entire
left hand side without using connection formulas at $x'_a$ or $x'_b$.
Once one has found $\psi'(x)$ sufficiently far to the right of $x'_b$
in this way, one can write the parabolic cylinder functions in
quasiclassical form which can then be extended in this form all the
way to $x=0$. The right-hand wavefunction can be treated in the same
way. This device leads to considerable savings in labor.

We now apply these ideas to our problem. In what follows, we will
denote quantities pertaining to the left hand solution or the left
hand side of the well by either a single prime or a suffix $-$, while
analogous right-hand quantities will carry a double prime or a $+$
suffix. We consider an energy $E$ as drawn in \fno{fig4}, which leads
to turning points at $m'_a$, $m'_b$, $m'_c$ on the left hand
side, and $m''_c$, $m''_b$, and $m''_a$ on the right. The
wavefunction $C'_m$ will decay away from the well bottom as
$m - m'_a$ decreases, oscillate in the classically allowed
region $m'_a < m < m'_b$ \cite{fn2,fn3}. In the region just to the
right of $m'_b$ it will consist of a decaying part and a growing part.
The new feature will be encountered at $m'_c$ where $E = U_*$. For
$m > m'_c$, both the growing and decaying parts will acquire oscillatory
envelopes. Similar remarks apply to the right side wavefunction $C''_m$.

\subsection{DPI wavefunction in the leftmost forbidden region}
\label{lefttt}
We are now ready to find the wavefunction explicitly. Let us
start constructing $C'_m$ from the left. The Hamilton-Jacobi equation
in \eno{Eandv} has the general solution
\bea
\cos q(m) &=& {-t_1(m) \pm [t^2_1(m) - 4t_2(m) f(m)]^{1/2}
              \over 4t_2(m)}; \label{cosq} \\
f(m) &=& w(m) - 2t_2(m) -E. \label{fofm}
\eea
This leads to four values of $q(m)$ for any $E$, since if $q$ is a
solution, so is $-q$. In the region $m \le m'_a$, since $E < U_0$,
all four solutions are pure imaginary. We write the two which lead
to decaying wavefunctions as $\mu - \mu'_a$ becomes large and negative
as
\beq
q_{1,2} = -i\kap_{1,2}(\mu), \label{q12}
\eeq
where $\kap_2 > \kap_1 > 0$, and the corresponding DPI solutions as
\bea
C'_{m,1} &=& A' |v_1(m)|^{-1/2} \exp\lp i\int^m q_1(m') dm'\rp, 
          \label{Cm1} \\
C'_{m,2} &=& B' |v_2(m)|^{-1/2}
              \exp\lp i\int_{m'_c}^m q_2(m') dm'\rp, \label{Cm2}
\eea
with $v_i(m) = v(q_i(m),m)$. We take $A'$ and $B'$ to be real without
any loss of generality. Note that we have left the lower limit
of the phase integral for $C'_{m,1}$ unspecified and written it as
$m'_c$ for $C'_{m,2}$. The reasons for this will become clear shortly.

To see how these two solutions behave as $m$ approaches $m'_a$, and
continues beyond this point, let us note that
\beq
\cosh\kap_{1,2} = {|t_1| \mp [t^2_1 - 4t_2 f]^{1/2}
              \over 4t_2}. \label{chkap} 
\eeq
As $m \to m'_a-$, $\cosh\kap_1 \to 1$, i.e., $\kap_1 \to 0$,
while $\cosh\kap_2 \to -1 + 2 |t_1|/4t_2 > 1$. As we cross the
point $m'_a$, $q_1$ will become real, while $q_2$ will continue to
be pure imaginary and large. Thus the solution
$C'_{m,2}$ continues to hold at $m'_a$, while $C'_{m,1}$ breaks down
at $m'_a$ \cno{fn3}, and must be related
to a the solution for $m>m'_a$ by a connection formula. It is clear
that the wavevector(s) for $C'_{m,2}$ will continue to be given by
\eno{chkap} as $m \to m'_b-$, while those for $C'_{m,1}$ will again
approach zero, necessitating the use of connection formulas to go on
to $m > m'_b$. It is here that the technical remark about sidestepping
the use of connection formulas that was made in connection
with the continuum antisymmetric double-well is relevant. The solution
$C'_{m,1}$ can be approximated by a harmonic oscillator wavefunction
provided the energy $E$ is not very far from the minima of $U_0$.
The asymptotic forms of this wavefunction give us the quasiclassical
wavefunction in the regions $m < m'_a$ and $m'_b$ more simply. We
therefore turn to this subproblem.

\subsection{Jumping across the potential well}
\label{jump}

The assumption that $E - U_0(\mu_{0\pm})$ is not very large, means
that $q_1(m)$ is never very far from zero, and we may expand
$\hsc(q,m)$ in powers of $q$ and $m + m'_0$. As in I, we write
\beq
\hsc(q,m) = E_- + {1\over 2 M_-} q^2 +
           \hf  M_- \om_-^2 (m - m_{0-,})^2 + \cdots. \label{hharm}
\eeq
Since $q$ and $m$ are conjugate variables, we 
can write $C'_{m,1}$ as the solution to the differential
equation
\beq
\hsc(-i\ptl_m,m)C'_{m,1} = E C'_{m,1}. \label{hscdiff}
\eeq
Introducing two new variables $z$ and $\nu'$ by the equations
\bea
m &=& \mzm + (2M_- \omzm)^{-1/2}z, \label{defz}\\
E   &=& E_- + (\nu' + \tshf) \omzm, \label{defnu'}
\eea
within the approximation (\ref{hharm}), the differential equation
becomes that for the parabolic cylinder functions:
\beq
\left[ {d^2 \over dz^2} + \lp\nu' + \hf - {z^2 \over 4}\rp \right]
      C'_{m,1} = 0. \label{deqz}
\eeq
If we take as the two linearly independent solutions the standard
forms $D_{\nu'}(z)$ and $D_{\nu'}(-z)$ \cno{bod}, the former must be 
rejected as it diverges for $z \to -\infty$. We accordingly write
\beq
C'_{m,1} = A' (-1)^{\ell'} D_{\nu'}(-z), \label{CtoD}
\eeq
where $A'$ is the constant in \eno{Cm1}, and the additional factor
$(-1)^{\ell'}$, where we shall define $\ell'$ shortly, is another
constant introduced for later convenience.

As $z \to -\infty$, $D_{\nu'}(-z) \sim (-z)^{\nu'} e^{-z^2/4}$, and
one can show with a little work that [modulo the factor $(-1)^{\ell'}$]
this is indeed the DPI form for $C'_{m,1}$ with the approximation
(\ref{hharm}) \cno{fn4}. For $z \to +\infty$, on the other hand,
\bea
D_{\nu'}(-z) &\sim& \cos(\pi\nu') z^{\nu'} e^{-z^2/4}
               \lp 1 - {\nu'(\nu' - 1) \over 2z^2} + \cdots \rp \nnu \\
             &&\quad + {\sqrt{2\pi} \over \Gam(-\nu')}
                z^{-\nu' - 1} e^{z^2/4} 
               \lp 1 + {(\nu'+1)(\nu'+2) \over 2z^2} + \cdots \rp.
                 \label{Dasymp}
\eea
Note that this form has both decaying and growing components. In fact,
the latter component vanishes only if $\nu'$ is a positive
a integer. As explained earlier, it is essential for our DPI solution
$C'_m$ to contain a growing component. We therefore allow for its
presence by writing
\beq
\nu' = \ell' + {\eps'\over \omzm}, \label{defeps}
\eeq
where $\ell'$ is a positive integer, and $\eps'$ is a shift defined to
lie in the interval $(-1/2, 1/2)\omzm$. In fact, we expect $\eps'$
to be very close to zero for any state in which there is large probaility
of finding the particle in the left well. We can make the vanishing of
the growing component in $C'_{m,1}$ more manifest by writing
\beq
{1\over \Gam(-\nu')} = - {\sin(\pi\nu') \over \pi} \Gam(1 + \nu')
             \apx (-1)^{\ell' + 1}(\ell'!){\eps'\over\omzm}.
        \label{Gamin}
\eeq
Combining Eqs.~(\ref{CtoD})--(\ref{Gamin}), we thus find that for
$m$ beyond $m'_b$,
\beq
C'_{m,1} \apx A'\lp
            \cos{\pi\eps'\over \omzm} z^{\nu'} e^{-z^2/4}
          -\sqrt{2\pi}(\ell'!){\eps'\over \omzm}
                          z^{-\nu'-1}e^{z^2/4} \rp. \label{Cm1rt}
\eeq

\subsection{DPI form in ordinary forbidden region}
\label{ordforbid}

The next step is to write the solution for $C'_m$ in such a way that
it holds in the entire region $m'_b < m < m'_c$ \cno{fn3}. For $C'_{m,2}$,
\eno{Cm2} already meets this demand, since just as at the turning point
$m'_a$, $q_2$ stays imaginary and negative as $m$ passes through $m'_c$.
For $C'_{m,1}$, on the other hand, 
\eno{Cm1rt} only holds in a region where the parabolic approximation to
$U_0$ is good, and may not hold near $m'_c$. It is clear, however, that
the wavevector $q_1$ associated with $C'_{m,1}$ is again given by
$-i\kap_1$ with $\kap_1$ given by \eno{chkap}. Hence it must be
possible to write $C'_{m,1}$ in the DPI form
\beq
C'_{m,1} =  |v_1(m)|^{-1/2}\left[
              Q' \exp\lp -i\int_{m'_b}^m q_1(m') dm'\rp
              + R' \exp\lp i\int_{m'_b}^m q_1(m') dm'\rp
               \right], \label{Cm1dpi}
\eeq
where $Q'$ and $R'$ are coefficients which we expect to be proportional
to $A'$. To find these, let us first calculate the phase integral in
the parabolic approximation. Since $q_1 = 0$ at $m'_b$, we have
\beq
E = \hsc(0,m'_b) = \hsc(i\kap_1(m),m), \quad (m > m'_b). \label{defkap}
\eeq
Using \eno{hharm}, we get
\beq
\kap_1(m) \apx M_-\omzm \bigl[ (m-\mzm)^2 - (\mpb - \mzm)^2)^{1/2}
                         \bigr].  \label{kap1par}
\eeq
Therefore, if $(m-\mpb) \gg (\mpb - \mzm)$, we obtain
\bea
\exp\lp - \int_{\mpb}^m \kap_1(m') dm'\rp
   &=& \lp 2 {m - \mzm \over \mpb - \mzm} \rp^{\nph}
        e^{\tshf(\nph)} \nnu \\
     &&\ \times \exp\lp -\hf M_-\omzm (m-\mzm)^2 \rp,
    \label{pipar}
\eea
where we have used the fact that
\beq
\tshf M_-\omzm^2 (\mpb-\mzm)^2 = (\nph)\omzm. \label{Eattp}
\eeq 
Next we note that (a) from the definition of $z$, \eno{defz}, the last
exponential in \eno{pipar} is nothing but $\exp(-z^2/4)$, and that
(b) $|v_1(m)| \apx \kap_1(m)/M_- \apx \omzm (m-\mzm)$. Since
$m - \mzm \propto z$, it follows that the firat term in \eno{Cm1dpi}
varies as $z^{\nu'} \exp(-z^2/4)$, i.e., precisely as the first term in
\eno{Cm1rt}. Thus $Q'$ is indeed proportional to $A'$, and a little
algebra plus the use of \eno{Eattp} gives
\beq
Q' = \lp {\omzm \over 2 M_-} \rp^{1/4} e^{-\tshf(\nph)}
        (\nph)^{\tshf(\nph)} \cos{\pi\eps'\over \omzm} A'
   \equiv \al'A'. \label{alpr}
\eeq
In the same way, we can show that
\beq
R' = -\lp {\omzm \over 2 M_-} \rp^{1/4} e^{\tshf(\nph)}
          (\nph)^{-\tshf(\nph)} \sqrt{2\pi}(\ell'!)
              {\eps'\over \omzm} A'
                 \equiv \be' {\eps'\over\omzm} A'. \label{bepr}
\eeq
The definitions of the factors $\al'$ and $\be'$ which we have
introduced for later convenience can be read off these equations.
Note that since we took $A'$ to be real, $Q'$ and $R'$ are also
real.

To summarize where we are,
the complete DPI solution for $C'_m$ in the region $m'_b < m < m'_c$
is given by the sum of \etwo{Cm2}{Cm1dpi}. The terms
in $B'$ and $R'$ are exponentially growing with increasing $m$, while
the term in $Q'$ is exponentially decreasing. The next step is to
connect this solution to the DPI form in the region $m'_c < m$.
As already stated, the turning point $m'_c$ is the irregular one
under the barrier, where exponentially growing and decaying
wavefunctions acquire oscillatory envelopes as it is crossed. 

Before we use the connection formulas at $m'_c$, it is useful to see
how the quasiclassical wavevector behaves near this point.
Since $\cos q(m'_c) = \cos q^* = -t_1(m'_c)/4t_2(m'_c)$, it follows
that the discriminant in \eno{cosq} vanishes at $m'_c$, and {\it both}
$q_1$ and $q_2$ tend to the same value $-i\kap'_c$, where
\beq
\cosh\kap'_c = (|t_1|/4t_2)_{m=m'_c}. \label{kapprc}
\eeq
As we cross $m'_c$, the discriminant in \eno{cosq} becomes negative
and $\cos q(m)$ [and therefore $q(m)$] becomes complex. We separate
$q(m)$ into its real and imaginary parts, and write two distinct
solutions as
\beq
q_{d,g}(m) = \pm i\kap(m) + \chi(m), \quad (m>m'_c) \label{kapchi}
\eeq
where both $\kap$ and $\chi$ are real and positive. The subscripts
$d$ and $g$ stand for `decaying' and `growing'.

\subsection{DPI form in oscillatory forbidden region}
\label{oscforbid}

The connection formulas to be used at $m'_c$ were derived in
Sec.~IV of \rno{jmp2}. The parts of $C'_m$ multiplying $B'$,
$Q'$, and $R'$ correspond respectively to the cases there
labelled $(\sig_1,\sig_2) = (+1,-1)$, $(-1,+1)$, and $(-1,-1)$.
The $B'$ part is given by
\beq
C'_{m,2} = B' \left[ {1\over \sqrt{s_g(m)}}
            \exp\lp i\int_{m'_c}^m q_g(m') dm' - {\pi \over 2}\rp
                  + {{\rm c.c.}} \right],
           \label{Cm2mid}
\eeq
where
\bea
s_g(m) &=& 8t_2(m) \sinh\kap(m) \sin\chi(m) \sin q_g(m) \nnu \\
       &=& 8t_2 \sinh\kap \sin\chi
            (\sin\chi\cosh\kap - i \cos\chi\sinh\kap). \label{defsg}
\eea
Likewise, the coefficient of $Q'$, which we call part 1a, connects
to
\beq
C'_{m,1a} = e^{-\Gam'} Q' \left[ {1\over \sqrt{s_d(m)}}
            \exp\lp i\int_{m'_c}^m q_d(m') dm' \rp
                  + {{\rm c.c.}} \right],
           \label{Cm1a}
\eeq
where
\beq
s_d(m) = 8t_2(m) \sinh\kap(m) \sin\chi(m) \sin q_d(m) = [s_g(m)]^*,
            \label{defsd}
\eeq
and $\Gam'$ is the phase integral,
\beq
\Gam' = \int_{\mpb}^{m'_c} \kap(m') dm', \label{Gampr}
\eeq
which we acquire in changing the lower limits of the $m$ integrals
from $m'_b$ to $m'_c$.
Lastly, the term in $R'$, which we call part 1b, connects to
\beq
C'_{m,1b} = \hf e^{\Gam'} R' \left[ {1\over \sqrt{s_g(m)}}
            \exp\lp i\int_{m'_c}^m q_g(m') dm' \rp
                  + {{\rm c.c.}} \right].
           \label{Cm1b}
\eeq

Equations (\ref{Cm2mid}), (\ref{Cm1a}), and (\ref{Cm1b}), give us the
complete wavefunction $C'_m$ in the central region near $m=0$. To
simplify the writing, we denote
\beq
\Phi'_{\lam_1\lam_2}(m) =
         \int_{m'_c}^m [\lam_1\kap(m') + i\lam_2\chi(m')]dm',
             \label{Phill}
\eeq
where $\lam_1$ and $\lam_2$ can be $\pm 1$ independently. In other
words, the subscripts on $\Phi'$ give the signs of the real and
imaginary parts [$\Phi'_{++}$ is the integral of $(\kap+i\chi)$,
$\Phi'_{-+}$ that of $(-\kap+i\chi)$, etc.]. The complete DPI solution
for $C'_m$ can then be written as
\beq
C'_m = \left[ e^{-\Gam'} Q' {e^{\Phi'_{-+}(m)}\over \sqrt{s_g^*(m)}}
          + \lp \tshf e^{\Gam'} R' - i B' \rp
         {e^{\Phi'_{++}(m)}\over \sqrt{s_g(m)}}\right] + {{\rm c.c.}}
   \label{Cpm}
\eeq

The wavefunction from the right, $C''_m$ can now be written down at
once. We define quantities with double primes in exact correspondence
with those for $C'_m$. The analog of \eno{Cpm} is then
\beq
C''_m = \left[ e^{-\Gam''} Q'' {e^{\Phi''_{-+}(m)}\over \sqrt{s_g^*(m)}}
          + \lp \tshf e^{\Gam''} R'' - i B'' \rp
         {e^{\Phi''_{++}(m)}\over \sqrt{s_g(m)}}\right] + {{\rm c.c.}}
   \label{Cdpm}
\eeq
The only issue requiring any thought is what the sign suffixes in $\Phi''$
should mean. By defining a new variable $n = -m$, so that the problem
for $C''_m$ becomes completely isomorphic to that for $C'_m$, and
then transforming back to $m$, one can show that
\bea
\Gam'' &=& \int_{m''_c}^{m''_b}\kap(m') dm', \label{Gamdp} \\
\Phi''_{-+}(m) &=& 
         \int_m^{m''_c}[-\kap(m') + i\chi(m')] dm', \label{Phidp}
\eea
etc. Note that since $m < m''_c$ in the center, and
$m''_c < m''_b$, these integrals are
written so that the lower limit is less than the upper limit. Thus
the suffixes on $\Phi''$ give the true signs of its real and imaginary
parts.

\subsection{Matching of left and right wavefunctions}
\label{match}

It remains to see if \etwo{Cpm}{Cdpm} are the
same function. We note that $\Phi''_{-+}(m)$ has the same integrand as
$\Phi'_{+-}(m)$ if $m$ is taken to be the upper limit for both
integrals. We further note that the remaining $m$ dependence in both
terms is $(s^*_g)^{-1/2}$. Similar remarks apply to the $\Phi'_{-+}$
and $\Phi''_{+-}$ terms. Thus, we conclude that $C'_m$ will equal
$C''_m$ if the following conditions are obeyed:
\bea
e^{-\Gam''} Q'' e^{\Phi''_{-+}(m)} &=& 
    \lp \tshf e^{\Gam'} R' + i B' \rp e^{\Phi'_{+-}(m)}, \label{cond1} \\
e^{-\Gam'} Q' e^{\Phi'_{-+}(m)} &=& 
    \lp \tshf e^{\Gam''} R'' + i B'' \rp e^{\Phi''_{+-}(m)}. \label{cond2}
\eea
To simplify these conditions we note that
\bea
\Phi'_{+-}(m) - \Phi''_{-+}(m) &=&
       \int_{m'_c}^{m''_c} [\kap(m) - i\chi(m)] \,dm, \nnu \\
   &\equiv& \Gam_c - i\Lam_c, \label{defGLc}
\eea
where $\Gam_c$ and $\Lam_c$ are the real and imaginary parts of the
integral, and the subscript `c' indicates that the integrals extend over
the central region of $m$. Equations (\ref{cond1}) and (\ref{cond2})
can now be written as
\bea
Q'' &=& \lp \tshf e^{\Gam'} R' + i B' \rp
       e^{\Gam''} e^{\Gam_c - i\Lam_c}, \label{cond3} \\
Q' &=& \lp \tshf e^{\Gam''} R'' + i B'' \rp
       e^{\Gam'} e^{\Gam_c - i\Lam_c}. \label{cond4}
\eea
If we recall that [see \etwo{alpr}{bepr}] $Q'$
and $R'$ are proportional to $A'$, and likewise for $Q''$ and $R''$,
these equations are two complex equations in the four real quantities
$A'$, $B'$, $A''$ and $B''$. To solve them we first note that the
imaginary parts on the right hand sides must vanish. This yields
\bea
B' &=& \tshf R' e^{\Gam'} \tan\Lam_c, \label{solBp} \\
B'' &=& \tshf R'' e^{\Gam''} \tan\Lam_c. \label{solBdp}
\eea
Substituting these in \etwo{cond3}{cond4}, we
obtain after some simplification
\bea
R' = 2 e^{-\Gam_G} \cos\Lam_c \,Q'', \label{RQ1} \\
R'' = 2 e^{-\Gam_G} \cos\Lam_c \,Q', \label{RQ2}
\eea
where $\Gam_G$ is the total Gamow factor
\beq
\Gam_G = \int_{m'_b}^{m''_b} \kap_1(m) dm, \label{Gamow}
\eeq
and the subscript `1' on $\kap$ is to remind us that we must
use the imaginary part of the wavevector that goes to zero at the
turning points $m'_b$ and $m''_b$.

\subsection{The eigenvalue condition, and the diabolo}
\label{split}

The simplest way of solving \etwo{RQ1}{RQ2} is
terms of the ratios $\al'$ and $\be'$ defined in \etwo{alpr}{bepr},
and the analogous ratios $\al''$ and $\be''$.
Equating the products of the left hand and right hand sides, and
simplifying a little, we get
\beq
\eps'\eps''  = 4 {\al'\al'' \over \be'\be''}
               \omzm\omzp \cos^2\Lam_c e^{-2\Gam_G}. \label{eval}
\eeq
This is our eigenvalue condition. To understand it better,
we first note that the right hand side is exponentially small on
account of the square of the Gamow factor $e^{-\Gam_G}$.
Thus, ignoring for the moment the possibility that $\cos\Lam_c$ may
vanish, either both $\eps'$ and $\eps''$ must be of order
$e^{-\Gam_G}$, or at the other extreme, one must be of order
unity, and the other of order $e^{-2\Gam_G}$. Suppose that
$\eps' = O(e^{-2\Gam_G})$, and $\eps'' = O(1)$. Let us take $A' = 1$.
Then from \etwo{alpr}{bepr}, we see that $Q' = O(1)$, while
$R' = O(e^{-2\Gam_G})$. It then follows from \etwo{RQ1}{RQ2}
that $Q'' \sim R'' = O(e^{-\Gam_G})$, and in turn from the
double primed analogs of \etwo{alpr}{bepr} that
$A'' = O(e^{-\Gam_G})$. Lastly, since $\Gam'' < \Gam_G$,
\eno{solBdp} implies that $B''$ is also exponentially small
\cno{fn5}. Hence, it follows that the entire wavefunction on the
right hand side of the well, $C''_m$ is exponentially small compared
to the left hand part $C'_m$. In other words, there is negligible
mixing of the states in the left and right hand well. This is exactly
what we expect when the energies of the two states in the absence of
tunneling differ by much more than the tunneling matrix element
itself.

The more interesting case, therefore, is that in which both $\eps'$
and $\eps''$ are of order $e^{-\Gam_G}$. In the defining equations for
the $\al$'s and $\be$'s, \etwo{alpr}{bepr}, we can then neglect
the $\eps$'s to very good approximation. This yields
\bea
{\al' \over \be'} &\approx&  - {1\over \sqrt{2\pi}\ell'!}
                    \lp \lph \rp^{\lph} e^{-(\lph)}, \nnu \\
         &=& -{g_{\ell'} \over 2\pi}, \label{albybe}
\eea
where $g_n$ is the standard curvature correction \cite{wf,amjp}
in the phase integral expression for the tunnel splitting
(see e.g., Eq.~(4.10) of paper I). This quantity tends to 1 rapidly
as $n$ gets large:
$g_0 = (\pi/e)^{1/2} \approx 1.075$, $g_1 \approx 1.028$,
$g_2 \approx 1.017$, $\ldots$.
The right hand side of \eno{eval} can thus be written as
$\Dta^2(\ell',\ell'')/4$, where
\beq
\Dll = {2 \over \pi} (g_{\ell'}g_{\ell''})^{1/2}
                      (\omzm\omzp)^{1/2} e^{-\Gam_G}\cos\Lam_c.
    \label{defDta}
\eeq
We further define
\bea
\eps &=& \tshf(\eps'+\eps'')
          = E - \hf\lp E_- + E_+ + (\ell'+\tshf)\omzm
                     +(\ell''+\tshf)\omzp \rp, \label{Eavg} \\
\dta &=& \eps'' - \eps'
          = \lp E_- - E_+ + (\ell'+\tshf)\omzm
                     - (\ell''+\tshf)\omzp \rp. \label{Edif}
\eea
With these definitions, \eno{eval} can be rewritten as
\beq
\eps = \pm \hf [\dta^2 + \Dta^2(\ell',\ell'') ]^{1/2}. \label{diab}
\eeq

Equation (\ref{diab}) is the complete formal solution to the problem
of tunneling in an asymmetric double well. Along with
Eqs.~(\ref{defDta})--(\ref{Edif}), it is the 
analog of the general phase integral formula for the tunnel
splitting in a symmetric double well that we found in I [see
Eq.~(4.38) there]. Since there is no great need for having the
final answer in simple closed form for the specfic problem (\ref{ham}),
and since the general procedure is fully explained in Sec.~IV.E of
I, we do not bother to extract the singular $\ln J$ parts of the
$\Gam_G$ integral. 

It is immediately obvious that the eigenvalues in \eno{diab}
are what we would get from a two level system Hamiltonian
\beq
\ham_{\rm TLS} = \hf \lp
                 \begin{array}{cc}
                  \dta & \Dll \\
                  \Dll & -\dta   \end{array} \rp, \label{htls}
\eeq
which is of course, just what we would expect.
The quantity $\eps$ is the energy measured from a convenient
reference point, while $\dta$, which depends on the fields $h_x$,
$h_z$, and the quantum numbers $\ell'$ and $\ell''$ of the states
whose mixing is being examined, is the offset between these energy
levels in the absence of tunneling. Equation (\ref{defDta}) 
gives the tunneling amplitude between these levels when the offset
is small, i.e., when the two levels are in approximate degeneracy. 
Note that although this amplitude is defined even for relatively
large offsets---offsets comparable to the intrawell spacings
$\ompm$---and indeed is not very sensitive to the value of the
offset, the concept of tunneling is physically sensible and useful
only when the offset is comparable to or less than the amplitude
$\Dta$. If $\dta \gg \Dta$, we get $\eps \apx \pm \dta/2$, i.e.,
$\eps'' \apx \dta$, $\eps' \apx \Dta^2/\dta$,
or the other way around. Then by the argument given after \eno{eval},
the mixing between the wells is negligible.

One may wonder if the above conclusion does not invalidate the
entire calculation. After all, we defined the shifts $\eps'$
and $\eps''$ assuming the wells were parabolic. Surely the corrections
to the energies from cubic and higher order corrections to the
potential are far larger than $\Dta$. Since the offset between the
levels must be tuned with exponential sensitivity, should not we
know the energies of the levels before tunneling to the same
sensitivity? The answer is no. The reason can be seen from
\eno{Cm1rt}. As we have seen, the amplitude of the growing part of
the wavefunction plays a key role in the tunneling. From \eno{Cm1rt},
we see that this amplitude is only {\it linearly} dependent on $\eps'$. 
Hence, a small error in locating the absolute position of the level
has little effect on the computed value of $\Dta$. To say it another
way, even though we know that the bottoms of the wells must be tuned to
exponential accuracy to get significant mixing between the two wells, we
cannot and need not determine the center of gravity of the two levels,
$\eps$, to the same accuracy to determine the tunneling amolitude itself.
This feature is also present in the symmetric case studied in I, although
there it is not so apparent, since Herring's formula gives $\Dta$
directly without making reference to the absolute energy level.

Secondly, it should be noted that \eno{diab} is nothing but the
equation for the diabolo. The splitting vanishes only when $\dta$
and $\Dta$ both vanish, which furnish the two conditions required
to determine the diabolical point. Since both $\dta$ and $\Dta$ will
in general have linear terms in the deviation from the diabolical
point, the energy surface is a double cone as asserted earlier.

Thirdly, let us ask if we recover the
results of paper I in the symmetric case, i.e., when $h_z = 0$. Then
$E_+ = E_-$, $M_+ = M_-$, and $\omzm = \omzp$. The only sensible case
is $\ell' = \ell'' \equiv n$, so that $\eps' = \eps''$, and $\dta = 0$.
The splitting is (up to an irrelevant sign) $\Dta(n,n)$, which is
precisely the tunnel splitting $\Dta_n$ computed in paper I. In
addition, however, we now have more explicit information about the
wavefunction. Proceeding as before, we see that $Q' \sim A' \sim 1$,
$R' \sim e^{-\Gam_G}$, and $B' \sim e^{-(\Gam_G - \Gam')}$.
(The double primed quantities are equal to their single primed
counterparts.) The conclusion about $B'$ is totally consistent with the
approach in paper I, which was based on Herring's formula. There one
takes the wavefunctions as $(C_{m,d} \pm C_{-m,d})/\sqrt{2}$, where
$C_{m,d}$ is the wavefunction of a state localized in the left well,
and which decays away from that well {\it in both directions}. If we
equate $C_{m,d}$ with $C'_{m,1a}$ (and, therefore,
$C_{-m,d}$ with $C''_{m,1a}$) in the central region, then we see
that $B'$, which by \eno{Cm2} gives the magnitude of the {\it growing}
part of $C'_m$ at $m=m'_c$, is also the order of magnitude of
$C''_{m,1a}$, the {\it decaying} part of $C''_m$ at $m = m'_c$.

\subsection{What is the origin of the diabolical points?}
\label{origin}

Lastly, it is extremely instructive to examine the problem when
$\Dta = 0$, i.e., $\cos\Lam_c = 0$, {\it without} necessarily imposing
the condition $\dta = 0$, for this gives insight into what causes
the quenching of the tunnel splitting.
Taking the imaginary parts of \etwo{cond3}{cond4} we see directly
that we must have $R' = R'' =0$, and that
\beq
Q'' = \pm B'e^{\Gam_c + \Gam''}, \quad
Q' = \pm B''e^{\Gam_c + \Gam'}.    \label{QtoB}
\eeq
Going back to \eno{bepr}, we see that $R'= 0$ requires either
$\eps' = 0$ or $A' = 0$, and likewise for $R''$, $\eps''$, and $A''$.
If $\dta\not= 0$, then both $\eps'$ and $\eps''$ can not vanish, 
and the only solution is $\eps' \not= 0$, $\eps'' = 0$,
$A'' = Q'' = B' = 0$, (or the one obtained by interchanging single and
double primes.) The only non-zero coefficients are $A'$, $Q'$, and
$B''$. From \etwo{alpr}{QtoB}, we see that only one of these
coefficients is independent, which can only be fixed by normalization.
Thus, we see that the part $C''_{m,2}$ proportional to $B''$ should
really be regarded as the extreme right-hand tail of a state localized
in the left well. If we denote this state by $\ket{L}$, and the
wavefunction $\tran{m}{L}$ by $C_L(m)$, then \cno{fn3},
\beq
C_L(m) = \cases{
     A' |v_1(m)|^{-1/2} \exp\lp i\int^m q_1(m') dm'\rp,
                & $ m < m'_a$, \cr 
     A' (-1)^{\ell'} D_{\ell'}(-z),
                & $m\apx m'_a\ {{\rm to}}\ m \apx m'_b$, \cr
      Q' |v_1(m)|^{-1/2}  \exp\lp -i\int_{m'_b}^m q_1(m') dm' \rp,
                & $m'_b < m < m''_b$, \cr
     B'' |v_1(m)|^{-1/2}
              \exp\lp -i\int_{m''_b}^m q_1(m') dm'\rp,
                & $m''_b < m$. \cr}
          \label{lwf}
\eeq
[Note that in the second line, we wrote $D_{\ell'}$, not $D_{\nu'}$,
and that in the third line, we have not bothered to
write the oscillatory exponential continuation in the region
$m'_c < m < m''_c$ correctly---see \eno{Cm1a}---since our aim now
is merely to indicate the general structure.] We can define a
right-hand function $C_R(m)$ analogously. Indeed it is now clear that
the two state Hamiltonian (\ref{htls}) is a truncation of the full 
Hamiltonian (\ref{ham}) in the $\ket L$, $\ket R$ basis.

The above argument shows that for any $H_z$ ($\dta \not= 0$), we
can tune $H_x$ so that $\Dta$ vanishes, at which point, the energy
eigenfunction is like $C_L(m)$ [or $C_R(m)$], which is localized in
one well and does not ``see" the other well at all! In ordinary
one dimensional quantum mechanics, this is of course impossible, since
a wavefunction like $C_L(m)$ which continues decaying with increasing
$m$ in the classically allowed region of the right hand well has the
wrong sign of the curvature in that well. In fact, this argument does
not depend on having $\dta\ne 0$.  If $\dta = 0$ in addition to
$\Dta = 0$, $C_L(m)$ and $C_R(m)$ are {\it independent} solutions of
Schr\"odingers equation, as is any linear combination, since they are
degenerate.

The above point of view helps
elucidate the origin of the quenching more clearly. Indeed, it is better
to think about the non-symmetric situation ($H_z \ne 0$) than the
symmetric one ($H_z = 0$). In continuum problems with a symmetric double
well, Herring's formula gives the splitting as proportional to
$[\psi_d(x)(d \psi'_d/dx)]_{x=0}$, where
$\psi_d(x)$ is a left-well localized wavefunction \cno{ll}. It is tempting to
think that the splitting in the spin problem vanishes
because the oscillatory envelope in $C_{m,d}$ in the central region
allows the discrete analog of $\psi_d(x)$ or $\psi'_d(x)$ to vanish at the
midpoint. This reasoning is false, as one can see from a close examination
of the symmetric case wavefunction, or even more clearly, by looking
at the situation when $\Dta = 0$ but $\dta \ne 0$. The condition $\Dta = 0$
can not be reduced to a {\it local} property of the wavefunction such as its
value or its slope at a particular point. Rather it is the {\it global}
property that the phase integral $\Lam_c$ be an
odd integer times $\pi/2$. From this perspective, the quenching is perhaps
better visualized as a manifestation of interfering Feynman trajectories
and the Berry phase, even though the value of this phase is more easily
found using the DPI method.

\section{Application to \Fe8}
\label{appfe8}

Let us now apply our general formalism to \Fe8. The problem of greatest
interest is the location of the diabolical points, and for that we need
only solve the conditions $\dta= \Dta = 0$. We have already given
formulas for the matrix elements and the critical curves in
Eqs.~(\ref{wm})--(\ref{Ustar}). The problem that remains is to
use these formulas to find $\dta$ and $\Dta(\ell', \ell'')$. To keep
the problem tractable and obtain answers in closed form, we will
assume that $h_z$ is small. Specifically, we will assume that the
reduced field $h_z$ defined above \eno{U0} is formally of order
$1/\baj$. This enables us to evaluate the turning points and action
integrals as expansions in powers of $h_z$. Also, it is convenient to
carry out all calculations in terms of the reduced variable $\mu$.

The first step is to obtain $\dta$. For this, we need to analyze the
critical curve $U_0(\mu)$. Its minima $\mu_{0\pm}$ are found to be
located at
\beq
\mu_{0\pm} = \pm \mu_0 + {h_z h_x^2 \over 1 - h_x^2} + O(h_z^2),
       \label{mu0pm}
\eeq
where
\beq
\mu_0 = (1 - h_x^2)^{1/2}. \label{mu0}
\eeq
The quantities $E_{\pm}$, $\ompm$, and $\Mpm$ defined through \eno{hharm}
are given by
\bea
E_{\pm} &\equiv& U_0(\mu_{0\pm})
        = - \left[ h_x^2 \pm 2\mu_0 h_z 
               + {h_z^2h_x^2 \over 1 - h_x^2} \right],
          \label{Epm} \\
\ompm &=& {2\lam^{1/2}\mu_0 \over \baj}
         \left[1 \pm {h_z \over 2\mu_0}
          \left( {1\over \lam} + {1 + 2h_x^2 \over 1 - h_x^2}\right)
             + O(h_z^2) \right], \label{ompm} \\
\Mpm &=& {1\over 2\lam h_x^2}
           \left[ 1 \mp {h_z \over \mu_0}
             \left({1\over \lam} - 2 \right) + O(h_z^2) \right].
                \label{Mpm}
\eea
Substituting these results in \eno{Edif}, we obtain
\beq
\dta(h_z,\ell',\ell'') = 4 \mu_0 h_z
        + {2 \sqrt{\lam} \mu_0 \over \baj} (\ell' - \ell'')
        -{\sqrt{\lam} h_z \over \baj}(\ell'+\ell'' + 1)c_1(h_x)
        + O(\baj^{-3}), \label{ansdta}
\eeq
where
\beq
c_1(h_x) = {1 - h_x^2 + \lam(1 + 2h_x^2)
             \over \lam (1-h_x^2)}. \label{defc1}
\eeq

The next step is to evaluate $\Dta(\ell',\ell'')$, or more precisely
$\Lam_c$, the imaginary part of the phase integral defined
in \eno{defGLc}, since that by itself locates the diabolical points.
This in turn requires expressions for $\chi(m)$ and the points
$m'_c$ and $m''_c$. To obtain $\chi(m)$, we return to the Hamilton-Jacobi
equation \eno{Eandv}, write $q = \kap + i\chi$, and separate the equation
into its real and imaginary parts. Eliminating $\kap(m)$ from the two
equations that result, we obtain a single equation for $\chi(m)$, which
can be written as
\bea
&&4 t_2(m) X^2 - g(m) X + {t_1^2(m) \over 4t_2(m)} = 0; \label{reim} \\
&& g(m) = w(m) + 2 t_2(m) - E, \label{defg}
\eea
where $X \equiv \cos \chi(m)$. [We can find the equation obeyed by
$\kap(m)$ similarly, and we discover that it is again \eno{reim} with
$X = \cosh \kap(m)$.]

What value of $E$ should we use in \eno{reim}? As stated earlier, the
value of $\Dta$ is relatively insensitive to small changes in the
absolute position of $E$. Thus we certainly needn't incorporate the
exponentially small shifts caused by the tunneling itself. Secondly,
the tunneling is relevant only when $\dta \sim \Dta$. Thus it suffices
to set both $\eps'$ and $\eps''$ to 0 in this part of the calculation.
Then, Eqs.~(\ref{Eavg}), (\ref{Epm}), and (\ref{ompm}), we get
\bea
E &=& \hf\lp E_- + E_+ + (\ell'+\tshf)\omzm
                     +(\ell''+\tshf)\omzp \rp \nnu \\
  &=& -h_x^2 + {\sqrt{\lam}\mu_0 \over \baj}(\ell' + \ell''+1)
             + O(\baj^{-2}). \label{Ellp}
\eea
Substituting this along with the formulas for $w(m)$, $t_1(m)$, and
$t_2(m)$ in \etwo{reim}{defg}, yields the following equation for $X$:
\bea
&&(1-\lam)(1-\mu^2)X^2 -[ 1+h_x^2 - \mu^2 - \zeta(\mu)]X
       +{h_x^2 \over 1-\lam} = 0; \label{eqX} \\
&& \zeta(\mu) = {\sqrt{\lam} \over \baj}
                 [\mu(\ell'' - \ell') + \mu_0 (\ell' + \ell'' +1)].
                   \label{zeta}
\eea
We have separated out the term $\zeta(\mu)$ in \eno{eqX} as it is
of order $\baj^{-1}$ relative to the other terms. We can solve the
quadratic equation and expand the result as a power series in $\zeta$.
Recalling that $X = \cos^2\chi$, we obtain
\beq
\cos\chi(\mu) = {h_x \over \sqrt{(1-\lam)(1-\mu^2)}}
           \left[ 1 + {\zeta(\mu) \over 2 (1-h_x^2 - \mu^2)}
                    + O(\zeta^2) \right]. \label{anschi}
\eeq
By setting $\cos\chi = 1$, we can determine the points $\mu'_c$
and $\mu''_c$, for which it is now more convenient to write
$\mu_{c\pm}$ instead. If the correction $\zeta$ were absent, it is
easy to see that these points would be at $\pm\mu_{c0}$, where
\beq
\mu_{c0} = [(1-\lam - h_x^2)/(1-\lam)]^{1/2}. \label{muc0}
\eeq
Now, with $\zeta \not = 0$, we can find $\mu_{c\pm}$ as a series
in $\baj^{-1}$. Up to leading corrections, we get
\beq
\mu_{c\pm} = \pm \mu_{c0}
             -{1 \over 2\sqrt{\lam}\mu_{c0}\baj}
              [\mu_{c0}(\ell'' - \ell')
                 \pm \mu_0 (\ell' + \ell'' +1)]. \label{mucpm}
\eeq

The phase integral $\Lam_c$ is given by
\beq
\Lam_c = \baj \int_{\mu_{c-}}^{\mu_{c+}} \chi(\mu) d\mu.
        \label{Lam2}
\eeq
We will only evaluate this accurate to terms of order $\baj^0$. 
Let us first consider the corrections entailed by replacing the
limits by $\pm \mu_{c0}$. It is not difficult to show that without
the $\zeta$ term in \eno{anschi},
\beq
\chi(\mu) \apx {\sqrt{1-\lam}\over h_x}
              (\mu_{c0}^2 - \mu^2)^{1/2}
                   \ {{\rm as}}\ \mu \to \mu_{c0}. \label{chisqrt}
\eeq
Inclusion of the $\zeta$ correction does not change the square root
approach to zero, and can not change the coefficient to leading order
in $\baj^{-1}$. From \eno{mucpm}, $\mu_{c\pm}$ differs from
$\pm\mu_{c0}$ by terms of order $\baj^{-1}$, so ignoring these shifts
in the limits of the integral causes an error of order $\baj^{-3/2}$
in the integral. Hence, we have 
\beq
\Lam_c \apx \baj \int_{-\mu_{c0}}^{\mu_{c0}} \chi(\mu) d\mu
              + O(\baj^{-1/2}).
        \label{Lam3}
\eeq
The error is smaller than $O(\baj^0)$, and so may be ignored. The
remaining integral may be done exactly as in paper I. We write the
solution to \eno{anschi} by $\chi = \chi_0 + \Dta\chi$, where $\chi_0$
is the solution when the $\zeta(\mu)$ correction is ignored, and
$\Dta\chi$ is the $O(\zeta)$ correction. The $\chi_0$ term can be
integrated by parts, and yields
\bea
\Lam_{c0} &=& 2\baj\int_0^{\mu_{c0}} \chi_0(\mu) d\mu \nnu \\
          &=& \pi\baj[ 1 - h_x(1-\lam)^{-1/2}]. \label{Lamc0}
\eea
The $\Dta\chi$ term yields
\bea
\Dta\Lam_c &=& -\baj\int_{-\mu_{c0}}^{\mu_{c0}}  
              {\sqrt{1-\mu^2_{c0}} \over \sqrt{\mu^2_{c0} - \mu^2}}
              {\zeta(\mu) \over 2 (1-h_x^2 - \mu^2)} d\mu \nnu \\
           &=& -(\ell' + \ell'' + 1){\pi \over 2}, \label{Lamc1}
\eea
where the result follows after an elementary integration by substitution.
Note that the first term in $\zeta(\mu)$ integrates to 0 as it is odd.
The total result for $\Lam_c$ is (reverting to unscaled variables)
\bea
\Lam_c &=& \Lam_{c0} + \Dta\Lam_c \nnu \\
       &=& {\pi \over 2}\left[
             2J - (\ell' + \ell'') -
               2J{H_x \over H_c \sqrt{1-\lam}} \right]. \label{ansLam}
\eea

It is now elementary to find the conditions for a diabolical point.
The offset $\dta$ becomes zero at a certain value of $h_z$, and
the tunneling amplitude $\Dta$ becomes zero when $\Lam_c$ is an odd
multiple of $\pi/2$. Adding arguments $\ell'$ and $\ell''$ to indicate
the level numbers becoming degenerate, the diabolicity conditions are,
\bea
{H_z(\ell',\ell'') \over H_c}
    &=& {\sqrt{\lam}(\ell'' - \ell') \over 2 J}
        \left[1 + {\sqrt{\lam}\over 4\mu_0 J}
           (\ell' + \ell'' + 1) c_1(h_x) + O(J^{-2}) \right],
                 \label{diabhz} \\
{H_x(\ell',\ell'') \over H_c}
   &=& {\sqrt{1-\lam} \over J}
        \left[ J - n - \tshf (\ell' + \ell'' + 1) \right],
              \quad n = 0, 1, \ldots, 2J - (\ell' + \ell'' + 1).
                 \label{diabhx}
\eea
Note that what appears in these equations is $J$, not $\baj$. Secondly,
the restrictions on the integer $n$ in \eno{diabhx} are obtained, as
explained in I, by demanding that $\Lam_c$ be positive.

We obtain the perfect centered rectangular lattice of diabolical points
introduced in \Slab{intro} if we ignore the $c_1$ term in \eno{diabhz},
and also the restriction $H_z/H_c \ll 1$ used to
perform the calculation. From the viewpoint of our DPI calculations,
this exactness is somewhat mysterious. However, by the duality argument
of \Slab{intro}, if formulas (\ref{diabhz})
and (\ref{diabhz}) are correct for small $H_z$
and large $H_x$, then they also yield diabolical points for large $H_z$
and small $H_x$. Of course if the former set of points corresponds to
low lying levels, i.e., small values of $\ell'$ and $\ell''$, to which
our analysis applies, the latter set of points corresponds to
rather highly excited levels, to which the analysis does not apply
{\it prima facie}. It is nevertheless surprising that the formulas
should fit together so neatly. It is also somewhat surprising that
the experimentally determined \Fe8 diabolical points [for the cases
$(\ell',\ell'') = (0,0)$, $(0,1)$, and $(0,2)$] should lie on a centered
rectangular structure so closely,
since, as is known, the value of the measured $H_x$ period on the
$H_z = 0$ line is almost 50\% different from that predicted by
\eno{diabhx}. This feature is at present understood only on the basis
of numerical diagonalization of the spin Hamiltonian including fourth order
terms in $\bJ$. An analytic approach to this aspect of the problem remains
for the future.

\acknowledgments
This work is supported by the NSF via grant number DMR-9616749.
I am indebted to Wolfgang Wernsdorfer and Jacques Villain for
useful discussions and correspondence about \Fe8.

\begin{figure}
\caption{Diabolical points for the \Fe8 model Hamiltonian (\ref{ham})
for (a) $J=7/2$, and (b) $J =4$, as per the perfect lattice hypothesis.
The numbers next to the light dashed lines show the numbers of levels
that are simultaneously degenerate at all diabolical points on that
line. By scaling the axes and reflecting about the heavy dashed line,
we obtain points that are dual to one another; points on the line are
self-dual. The pattern in the other three quadrants is obtained by
reflecting about the $H_x$ and $H_z$ axes.}
\label{dpts}
\end{figure}

\begin{figure}
\caption{Critical energy curves for the Hamiltonian (\ref{ham}).}
\label{fig2}
\end{figure}

\begin{figure}
\caption{Asymmetric double for a massive particle in one dimension,
showing the turning points for an energy $E$.}
\label{fig3}
\end{figure}

\begin{figure}
\caption{Analog of \fno{fig3} for the discrete case, showing only the left
hand well. The critical curve $U_{\pi}$ is also not shown, as it does not
determine any turning points for low lying levels.}
\label{fig4}
\end{figure}

\begin{table}[t]
\caption{Numerical test of perfect lattice hypothesis for $J=5$. All diabolical points
are obtainable from those listed here by duality and the symmetries $H_x \to -H_x$,
$H_z \to -H_z$. In columns 4 and 5, $i_x = 2 (J - n) - (\ell' + \ell'' + 1)$, and
$i_z = \ell'' - \ell'$. The next two columns give, for each value of
$\lam$, the differences $\dta_x = |i_x - 2J H_x/H_c \sqrt{1-\lam}|$,
and  $\dta_z = |i_z - 2J H_z/H_c \sqrt{\lam}|$, where $H_x$ and $H_z$ are the numerically
computed coordinates of the diabolical point. The numbers in parentheses give the power of
10 multiplying the number preceding. Where the error is given as 0.00, it is less than our
machine accuracy.}
\vspace{0.2cm}
\begin{center}
\begin{tabular}{c c c c c c c c c c c}
{}& {}& {}& {}& {}& \multicolumn{2}{c}{$\lam = 0.10$} & \multicolumn{2}{c}{$\lam = 0.25$}
                    & \multicolumn{2}{c}{$\lam = 0.40$} \\
 $\ell'$ & $\ell''$ & $n$ & $i_x$ & $i_z$ & $\dta_x$ & $\dta_z$
                                          & $\dta_x$ & $\dta_z$   & $\dta_x$ & $\dta_z$ \\  
\hline
   0 & 0 & 0   &  9 &  0 &  3.21(-11) &  1.04(-11) &  1.41(-10) &  5.93(-12) &  1.36(-11) &  6.60(-11) \\
   0 & 0 & 1   &  7 &  0 &  5.19(-11) &  1.79(-12) &  1.77(-11) &  9.70(-12) &  1.18(-11) &  2.89(-11) \\
   0 & 0 & 2   &  5 &  0 &  4.37(-11) &  1.10(-12) &  1.38(-11) &  2.90(-11) &  8.12(-11) &  7.08(-12) \\
   0 & 0 & 3   &  3 &  0 &  1.85(-11) &  8.91(-14) &  2.12(-11) &  8.77(-13) &  1.41(-11) &  5.93(-13) \\
   0 & 0 & 4   &  1 &  0 &  5.91(-11) &  7.04(-14) &  3.84(-11) &  7.84(-13) &  1.66(-11) &  1.41(-13) \\
   0 & 1 & 0   &  8 &  1 &  1.27(-10) &  1.19(-11) &  5.86(-11) &  6.39(-11) &  6.60(-11) &  9.11(-12) \\
   0 & 1 & 1   &  6 &  1 &  3.21(-11) &  1.04(-11) &  8.12(-11) &  7.08(-12) &  5.17(-11) &  2.27(-12) \\
   0 & 1 & 2   &  4 &  1 &  6.65(-11) &  1.21(-12) &  3.49(-11) &  9.86(-12) &  3.88(-11) &  4.40(-11) \\
   0 & 1 & 3   &  2 &  1 &  3.84(-11) &  7.83(-13) &  1.11(-10) &  4.88(-12) &  1.18(-11) &  1.22(-11) \\
   0 & 2 & 0   &  7 &  2 &  1.52(-11) &  1.70(-11) &  2.31(-11) &  4.20(-11) &  7.31(-12) &  8.85(-11) \\
   0 & 2 & 1   &  5 &  2 &  1.79(-10) &  3.28(-11) &  2.67(-11) &  2.20(-11) &  2.88(-11) &  9.42(-11) \\
   0 & 2 & 2   &  3 &  2 &  1.11(-10) &  4.88(-12) &  1.67(-11) &  1.77(-11) &  1.31(-11) &  4.22(-11) \\
   0 & 3 & 0   &  6 &  3 &  3.16(-11) &  3.03(-11) &  2.99(-11) &  9.90(-11) &  1.92(-11) &  2.98(-11) \\
   0 & 3 & 1   &  4 &  3 &  4.18(-11) &  8.14(-12) &  1.53(-12) &  5.76(-11) &  2.58(-11) &  1.16(-11) \\
   0 & 4 & 0   &  5 &  4 &  3.55(-12) &  4.34(-11) &  4.38(-11) &  1.46(-10) &  3.32(-12) &  9.89(-11) \\
   1 & 1 & 0   &  7 &  0 &  6.64(-11) &  1.37(-11) &  3.17(-11) &  2.86(-11) &  6.87(-12) &  1.49(-10) \\
   1 & 1 & 1   &  5 &  0 &  7.16(-11) &  1.21(-11) &  3.16(-11) &  3.03(-11) &  1.39(-11) &  9.24(-11) \\
   1 & 1 & 2   &  3 &  0 &  5.31(-11) &  1.08(-12) &  2.63(-11) &  8.57(-12) &  1.80(-11) &  3.62(-12) \\
   1 & 1 & 3   &  1 &  0 &  1.11(-10) &  4.88(-12) &  8.94(-12) &  8.09(-12) &  2.65(-12) &  7.27(-12) \\
   1 & 2 & 0   &  6 &  1 &  3.91(-11) &  1.12(-11) &  4.93(-11) &  8.39(-11) &  6.40(-12) &  2.86(-10) \\
   1 & 2 & 1   &  4 &  1 &  3.41(-13) &  2.35(-11) &  5.11(-11) &  3.35(-11) &  9.35(-12) &  2.02(-10) \\
   1 & 2 & 2   &  2 &  1 &  2.30(-11) &  2.37(-12) &  2.24(-11) &  3.31(-11) &  4.48(-12) &  8.81(-12) \\
   1 & 3 & 0   &  5 &  2 &  3.20(-11) &  3.25(-11) &  1.91(-11) &  2.05(-10) &  3.47(-12) &  1.50(-10) \\
   1 & 3 & 1   &  3 &  2 &  2.83(-11) &  2.37(-11) &  2.00(-11) &  1.34(-10) &  1.21(-12) &  1.18(-10) \\
   1 & 4 & 0   &  4 &  3 &  2.09(-11) &  5.37(-11) &  6.41(-12) &  1.50(-10) &  1.19(-13) &  1.99(-10) \\
   2 & 2 & 0   &  5 &  0 &  2.33(-11) &  2.56(-11) &  1.79(-12) &  8.70(-11) &  2.70(-12) &  1.49(-10) \\
   2 & 2 & 1   &  3 &  0 &  2.28(-11) &  1.14(-11) &  2.33(-12) &  2.73(-11) &  7.50(-14) &  6.24(-11) \\
   2 & 2 & 2   &  1 &  0 &  6.55(-12) &  1.57(-13) &  5.61(-13) &  7.39(-12) &  6.28(-14) &  3.61(-11) \\
   2 & 3 & 0   &  4 &  1 &  1.15(-11) &  1.72(-11) &  4.41(-12) &  4.63(-11) &  4.33(-14) &  1.49(-10) \\
   2 & 3 & 1   &  2 &  1 &  1.31(-11) &  4.88(-11) &  2.15(-12) &  8.06(-11) &  1.58(-13) &  7.92(-11) \\
   2 & 4 & 0   &  3 &  2 &  9.02(-12) &  2.88(-10) &  2.19(-12) &  3.00(-11) &  3.96(-14) &  1.12(-10) \\
   3 & 3 & 0   &  3 &  0 &  4.21(-14) &  2.27(-11) &  5.38(-13) &  3.56(-11) &  1.46(-16) &  0.00      \\
   3 & 3 & 1   &  1 &  0 &  7.94(-13) &  1.29(-11) &  3.84(-13) &  6.54(-11) &  1.42(-14) &  2.39(-11) \\
   3 & 4 & 0   &  2 &  1 &  1.41(-12) &  5.26(-11) &  4.17(-15) &  4.45(-11) &  8.78(-16) &  1.49(-10) \\
   4 & 4 & 0   &  1 &  0 &  3.60(-14) &  3.22(-11) &  0.00      &  0.00      &  0.00      &  0.00      
\end{tabular}
\end{center}
\end{table}


\begin{references}
\bibitem{fn2} In fact, as we shall see, it will also have a small
component (denoted $C'_{m,2}$) that decays exponentially to the left.
\bibitem{fn3} As shown in Appendix A of \rno{jmp2}, all DPI solutions
break down in a narrow region of width $O(J^{1/3})$ surrounding a
turning point. Throughout this paper, when we give the region of
validity of such a solution in the form of an inequality such as
$m'_a < m$ or $m < m'_b$, it is to be tacitly understood that the
breakdown region is excluded. It would be very tedious to say so
explicitly every single time.
\bibitem{fn4} It is not necessary to find the precise choice of lower
limit in the phase integral in \eno{Cm1} which will lead to exact
agreement with \eno{CtoD}.
\bibitem{fn5} A similar argument applied to \eno{solBp}, along with
the estimate $R' = O(e^{-2\Gam_G})$, shows that $B'$, the amplitude
of the growing part of $C'_m$ at $m'_c$, is almost ``doubly" small,
i.e., of order $e^{-a\Gam_G}$, with $1< a < 2$.
\end{references}
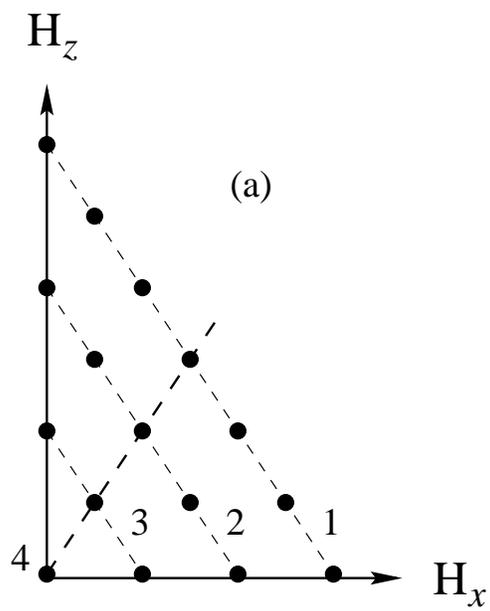
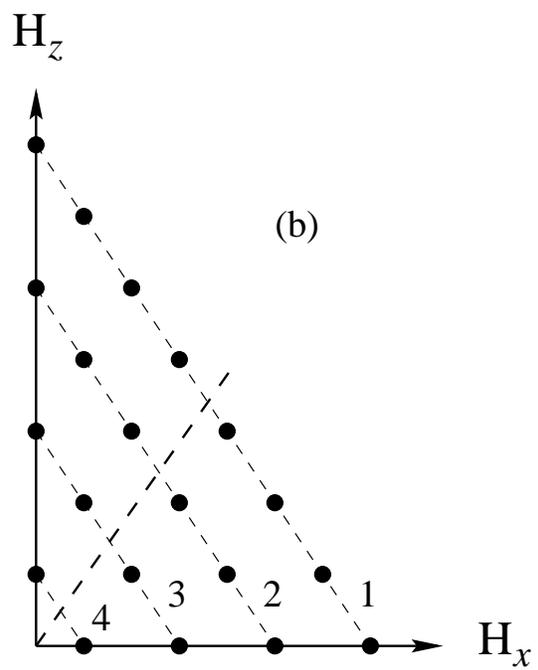
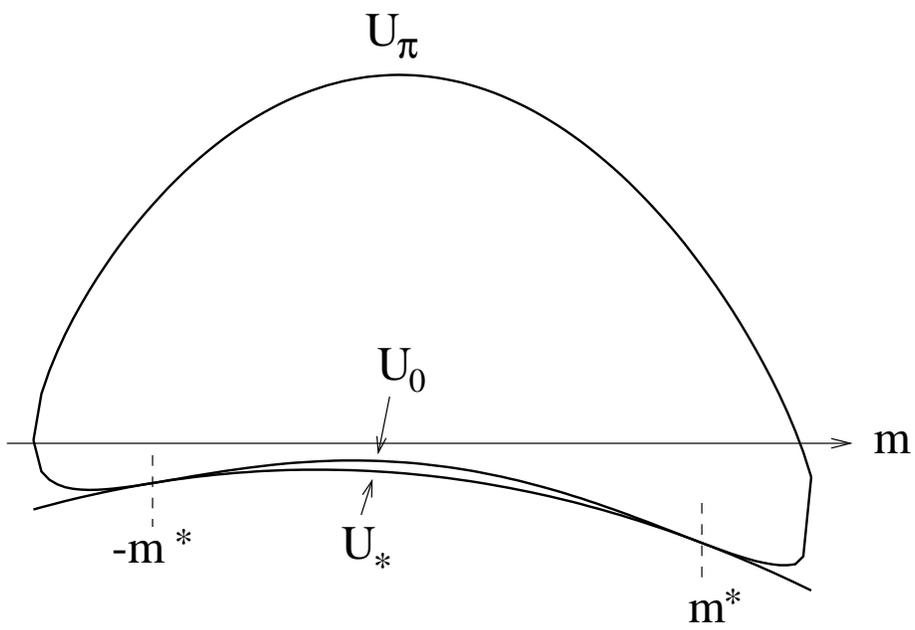
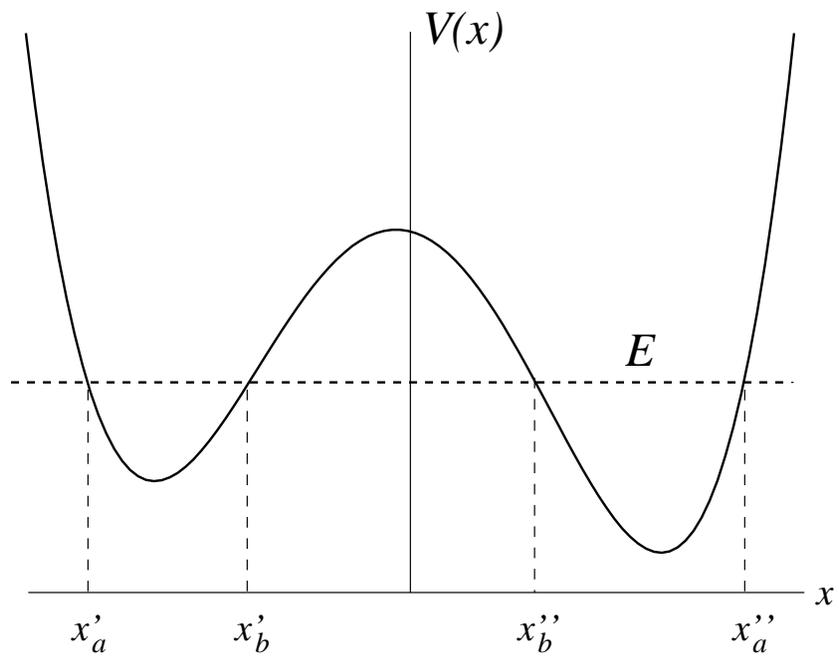
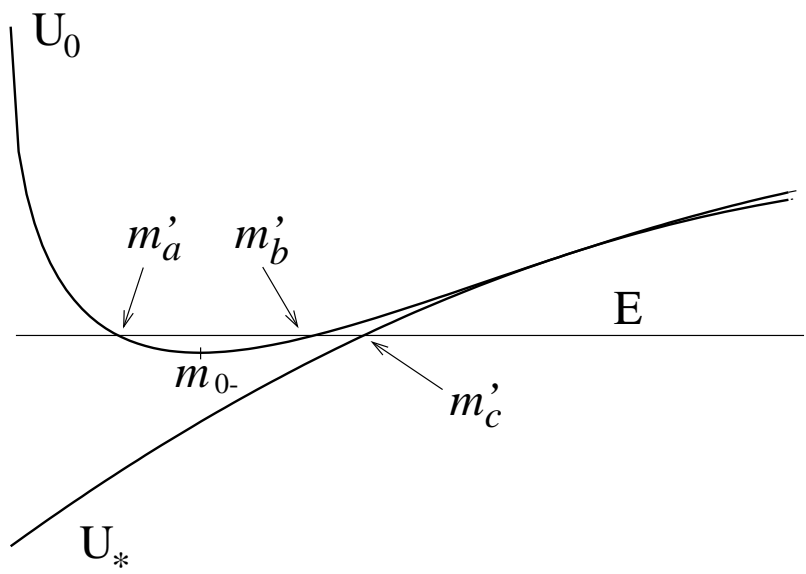
\end{document}